\documentclass[journal, a4paper]{IEEEtran}

\usepackage{graphicx}   
\usepackage{amsmath}    
\usepackage{textcomp}
\usepackage[noabbrev ]{cleveref}
\usepackage{siunitx}
\usepackage{bm}
\usepackage{tabularx}
\usepackage{array}
\usepackage[utf8]{inputenc}
\usepackage{tikz,pgfplots}
\usepackage[mode=buildnew]{standalone}
\pgfplotsset{compat=newest}
\usepackage{natbib}
\usepackage[draft]{todonotes} 
\newcolumntype{P}[1]{>{\centering\arraybackslash}p{#1}}

\usepackage{standalone}
\usetikzlibrary{calc, shapes}
\usetikzlibrary{patterns}

\newcommand{\NumberDen}{\rho_n}
\newcommand{\Source}{\mathbf{s}}
\newcommand{\Rec}{\mathbf{r}}
\newcommand{\Xp}{\mathbf{x_m}}
\newcommand{\X}{\mathbf{X}}

\newcommand{\TotCross}{\sigma_T}

\newcommand{\uu}{u_u(t)}
\newcommand{\up}{u_p(t)}
\newcommand{\Dvel}{\langle \tau \rangle}

\newcommand{\upi}{u_{p_j}(t)}

\newcommand{\upiTau}{u_{p_j}(t + t_s)}
\newcommand{\upin}{u_{p_{j-N}}(t)}
\newcommand{\upinSQ}{u_{p_{j-N}}^2(t)}
\newcommand{\upiSQ}{u_{p_j}^2(t)}
\newcommand{\wdwINT}{\int ^{t_{k}+t_{w}}_{t_{k}-t_{w}}}

\newcommand{\RmeasT}{d t_{pj-N}}

\newcommand{\Kfix}{K^{fix}}
\newcommand{\Kroll}{K^{roll}}
\newcommand{\KrollInt}{K^{roll}_{sum}}

\newcommand{\relVelTOF}{[\delta v/v]_{TOF}}
\newcommand{\relVelRoll}{[\delta v/v]^{roll}_{sum}}
\newcommand{\relVelFix}{[\delta v/v]^{fix}}

\begin{document}

    \title{Coda-Wave Monitoring of Continuously Evolving Material Properties and the Precursory Detection of Yielding}
    \author{Reuben Zotz-Wilson, Thijs Boerrigter, Auke Barnhoorn
    \thanks{}}
    \maketitle

  \begin{abstract}
The nominally incoherent coda of a scattered wavefield has been shown to be a remarkably sensitive quantitive monitoring tool. Its success is however often limited to applications where only moderate or localised changes in the scattering properties of the medium can be assumed. However, the compressional deformation of a relatively homogeneous rock matrix towards failure represents for a monitoring wavefield pronounced changes in both velocity and scattering power often due to a distribution of inelastic changes. We implement a rolling reference wavefield when applying Coda-Wave Interferometry and Coda-Wave-Decorrelation allowing relative velocity and material scattering power monitoring for such applications. We demonstrate how this modification enables the qualitative monitoring of stages in material deformation common to Unconfined Compressive Strength tests. In addition, the precursory/subtle onset of material yielding is identifiable in both the CWI and CWD methods, which was not possible when comparing to a fixed reference wavefield. It is therefore expected that this approach will enable these coda based methods to robustly monitor continuous, destructive processes at a variety of scales. Possible applications include critical infrastructure, landslide, and reservoir compaction monitoring where both the subtle continuous and sudden large changes in a material's scattering properties occur. 
\end{abstract}

\section{\label{sec:1} Introduction}
With the advent of affordable sensor networks, along with the ability to transmit large amounts of data wirelessly, the feasibility of evermore complex monitoring systems is increasing rapidly \citep{Brownjohn2007, Staehli2015}. Whether it be critical infrastructure structural health monitoring (bridges, dams, nuclear reactors) or environmental hazards (earthquakes, landslides, induced seismicity) or early warning detection to dynamic failure, continuous non-intrusive monitoring is crucial.\newline
Many works have demonstrated the remarkable sensitivity of the scattered wave to media-wide properties, through the estimation of velocity changes over fault zones \citep{Poupinet1984}, in volcanoes \citep{Ratdomopurbo1995, Matsumoto2001a, Gret2005, Sens-Schonfelder2006}, landslides \citep{Mainsant2012}, and within the lunar near surface \citep{Sens-Schonfelder2008}. Various ultrasonic laboratory experiments have shown the coda's sensitivity to changes at the mesoscopic scale in terms of stress, temperature, and water saturation \citep{Stahler2011,Zhang2012,Gret2006}. The Coda-Wave Interferometry (CWI) formulation presented by \citet{Snieder2002a} has served as the basis for many such relative time-lapse monitoring studies on solid media. In the field of optics Diffusing Wave Spectroscopy (DWS) \citep{Pine1990} has been used to study different aspects of strongly scattered light, and was later extended to acoustics as Diffusing Acoustic Wave Spectroscopy (DAWS) \citep{Page2000} relating phase changes in a wavefield to for example monitor the average displacement of scatterers in a fluidised suspension. \newline
In general the recent research applying CWI has focused on exploiting the sensitivity benefits of the coda-wave in monitoring subtle, often cyclical, non-destructive processes and thereby inferring the average velocity change. To the authors knowledge the only studies which have applied CWI well into a region of inelastic deformation have been limited to concrete structures either partially \citep{Zhang2016a} or fully under tensile forces \citep{Zhang2012}, of which only the latter shows the ability to track velocity changes close to the ultimate strength of the structure. \newline
The method termed Coda-Wave Decorrelation (CWD) as introduced by \citet{Larose2010} and later formalised by \citet{Rossetto2011} provides a spatiotemporal theoretical expression for the resulting decorrelation between a reference and perturbed wavefield due to the addition of a single or multiple localised scatterers \citep{Planes2015}. The ability of CWD to locate such a local change in scattering properties with the aid of a maximum likelihood inversion between the measured and theoretical decorrelation has  been shown in a laboratory setting \citep{Larose2010,Larose2015}, around an active volcano \citep{Obermann2013a} and most recently on a life-sized reinforced concrete structure \citep{Zhang2016a}. The focus of this branch of coda-wave studies is on relating the decorrelation coefficient $K$ between two wavefields to changes in material scattering properties. The most recent concrete structural health monitoring applications of CWD \citep{Larose2015,Zhang2016a} demonstrate its ability to identify the transition from elastic to inelastic deformation. However, there have been no published work which apply either CWI or CWD on a rigid material throughout the elastic and inelastic deformation up until catastrophic failure. Furthermore, in both works  macroscopic cracking is initiated under tensional bending where the stress field is localised, and therefore a more localised crack network results than would otherwise occur under a homogeneous tensional or compressional loading regime \citep{Paterson2005}. While such localised changes are well suited to the LOCADIFF algorithm \citep{Larose2010,Planes2013c}, where a more homogeneously distributed fracture network develops the resolution limit of the method will become an issue.\newline
In this study we explore the practical application of these coda-wave focused monitoring methods on a rock matrix undergoing continuous changes in bulk scattering power and intrinsic velocity due to a homogenous stress field with the goal of identifying defined stages of material deformation \citep{Heap2008} and precursory indication to material yielding. In particular, we show that for a fixed reference wavefield $\uu$ monitoring the onset of elastic deformation due to compressional loading, both CWI and CWD monitoring experience a rapid decay in sensitivity. As a result both are unable to identify the transition from elastic to inelastic deformation.\newline
In order to overcome this, a rolling reference wavefield $ \upin$ lagging $N$ measurements behind the most recently acquired $j^{th}$ monitoring wavefield $\upi$ is employed in a normalised cross-correlation formulation. With this simple modification, we show how one is able to identify three defined phases of material deformation common to laboratory Unconfined Compressive Strength (UCS) tests.  From this segmentation we are able to determine for two lithologies and three samples, precursory indicators to a materials yield point in both the CWI and CWD trends. It should be mentioned that while the idea of changing the reference has been applied before \citep{Obermann2013a,Gret2006,Snieder2002b}, and briefly discussed in a rolling formulation in the thesis of \citet{Planes2013a}, its application to enable the continuous monitoring of a material's properties throughout deformation up until catastrophic failure remains unexplored.
\section{\label{sec:2} Fixed-Reference Monitoring of Destructive Processes}
	\subsection{The Theory of CWI for Monitoring Changes in Velocity}
The formulation of CWI as presented by \citet{Snieder2002a} rests upon the understanding that a recorded wavefield $u_{u}(t)$ which has interacted with an unperturbed scattering medium can be represented as the summation of all possible paths $P$ through that medium as,
		\begin{equation}
			u_{u}(t) = \sum_P A_{P}(t),
        \label{eq:U_u}
	    \end{equation}
where $A_P$ denotes the wave which propagates along path $P$. The first major assumption concerning the medium itself is that each individual scatterer has stationary properties, therefore preventing a change in its scattering cross-section (size, shape, density and velocity). Additionally, it is assumed that the mean free path $l$, which is indirectly related to the averaged distance between scatterers is much greater than the dominant wavelength $\lambda$.

A perturbed wavefield $u_{p}(t)$ which has experienced a subtle change $\delta<<l$ in either the location of scatterers, the location of the source or the background medium velocity, can then be represented as,
  		\begin{equation}
			u_p(t) = \sum_{P} A_{P}(t- \tau_{P}),
        \label{eq:U_p}
	    \end{equation}
  where $\tau_P$ represents the travel time change along path $P$. This formulation therefore implies that provided the perturbation does not change the dispersion of the wavefield; only a change in the arrival time of $u_{u}(t)$ will occur. The Cross-correlation Coefficient $CC(t_{s})$ for a particular window of width $2t_w$ centred at time $t_k$ within the coda, $(t_k-t_w)$ to $(t_k+t_w)$ is often presented as,
		\begin{equation}
  CC( t_{s}) = \dfrac { \int ^{t_{k}+t_{w}}_{t_{k}-t_{w} }u_{u}( t) u_{p} ( t+t_{s})dt} { \Big[ \wdwINT u_{u}^{2}(t)dt \wdwINT u_{p}^{2}(t)dt \Big] ^{1/2}},
        \label{eq:CC}
	    \end{equation}
where $\uu$ and $\up$ are the unperturbed and perturbed waveforms respectively. This time shifted Cross-correlation Coefficient $CC(t_{s})$  will reach a maximum when the average travel time perturbation $\Dvel$ across all perturbed paths P is
		\begin{equation}
			 \Dvel = t_s 
		\end{equation}

A homogeneous relative velocity perturbation can then be determined to a first order approximation by,
	\begin{equation}
	\dfrac{\delta v}{v} = -\dfrac{\Dvel}{t_k},
	\label{eq:relVel}
	\end{equation}
	 If we now consider this in terms of a rock matrix the first assumption of CWI implies that the size, width and number of inhomogeneities (e.g. fractures, pores) remains constant in time. Furthermore, it requires that for a change in the phase to dominate the average distance between such inhomogeneities (the mean free path) is considerably larger than a wavelength. The constraint of a non-dispersive perturbation requires that changes in properties such as scattering or intrinsic attenuation are negligible. Provided only a subtle, cyclical, elastic deformation is applied to a medium with a sufficiently high source frequency, these assumptions generally hold within the time period between the perturbed and unperturbed wavefield. This is demonstrated for a rock matrix in the work by \citet{Gret2006} where CWI derived velocity changes are observed in a sandstone core sample due to a subtle elastic increase in uniaxial compressive stress.
	\subsection{Applying a CWD Approach to Monitoring Changes in Scattering Properties}
In order to enable the coda based monitoring of scenarios where inelastic material deformation occurs one can focus on the maximum cross-correlation coefficient $CC(t_s)$, instead of the $ t_{s}= \Dvel $ at which its maximum is found.  This is equivalent to the Coda-Wave Decorrelation methods discussed earlier, though with the distinction that here the main goal is not to find a single \citep{Larose2010} or several localised scatterer perturbations \citep{Planes2015} within the medium but to monitor a global perturbation within the region sampled by the coda-wave. 

	The scattering power of a medium can be defined in terms of its total total scattering coefficient $g_0$ \citep{Aki1975}. Assuming an idealised scattering media as a random distribution of $n$ point-like scatterers with number density $\rho_n$, within a background velocity $V_0$, and a propagating plane wave this coefficient can be defined as, 
	    \begin{equation}
  		g_{0} = \NumberDen{} \TotCross{} \equiv l^{-1},
        \label{eq:g0}
	    \end{equation}
	   which is inversely proportional to a materials mean free path $l$. By applying diffusive propagation theory, \citet{Rossetto2011} derived an expression for the theoretical decorrelation caused by localised  perturbation of the total scattering cross-section $\TotCross$. This requires the knowledge of a sensitivity kernel $Q( \Source{} ,\Xp{}, \Rec{} ,t)$ between a source $\Source{}$ and receiver $\Rec{}$ for the perturbation location $\Xp{}$ and a time $t$ within the coda.
	Re-writing this in terms of a change in a materials total scattering coefficient $g_0$ with a background velocity $V_{0}$, explicitly in terms of unperturbed and perturbed medium states gives, 
	\begin{equation}
		K^{T}(\X,t) = \dfrac{V_0}{2} \int_{V_{Q}} \left| \Delta g_{0_{p-u}}(x)  \right| Q( \Source{} ,x, \Rec{} ,t ) d V_Q(x),
		\label{eq:K_the_homo}
	\end{equation}
where $\X$ defines the ensemble of $n$ randomly distributed perturbation locations within the kernel volume $V_{Q}$, while $\left| \Delta g_{0_{p-u}}  \right(x)|$ locally defines the difference in total scattering coefficient between the perturbed and unperturbed material states. 
Considering \cref{eq:K_the_homo} in terms of the evolution of a materials scattering properties, any change in the size or impedance contrast of the scatterers will result in an increase in the decorrelation, as will the addition or removal of scatterers as the number density $\NumberDen{}$ changes. Furthermore, these changes will alter the intensity of the diffusive wavefield \citep{Paasschens1997,Pacheco2005} through the materials diffusion and attenuation coefficients, resulting in a change in the sensitivity kernel.

Equation \ref{eq:K_the_homo} is therefore able to describe both a change in the number of scatterers, through $\NumberDen$ or in the size and impedance contrast of scatterers, through $\TotCross$, provided one can assume the same sensitivity kernel $Q(\Source{} ,\X, \Rec{} ,t )$ in both states. For a large perturbation in $g_0$, it will become difficult to maintain this approximation as the diffusion and attenuation coefficients between the two medium states begin to diverge. In terms of the measured data the decorrelation coefficient ($K( t_{s}) = 1- CC( t_{s})$) between two recorded time-series will approach one, at which point it will only represent a spurious correlation between two time-series. However, provided that a weak perturbation can be ensured in the time interval $t_{u} - t_{p}$, the measured $K(t_{s})$ can be related to the modulus of the change in a materials scattering coefficient weighted by the sensitivity kernel.

\section{\label{sec:3} Rolling-Reference Coda Monitoring of Destructive Processes}
With the goal of ensuring the assumptions of CWI and CWD are satisfied throughout the long term monitoring of continuously evolving material scattering properties, we propose the use a rolling reference waveform when determining both the decorrelation coefficient and the relative velocity change. While the idea of manually selecting a different reference wavefield during monitoring is not new \citep{Gret2006, Obermann2013}, the uses of a rolling reference wavefield enabling the continuous monitoring of a material throughout deformation up until catastrophic failure remains unexplored. With regard to \cref{eq:CC} such a modification requires the fixed reference wavefield $\uu$ to be replaced by a rolling reference wavefield $ \upin$, in a monitoring sense lagging behind the most recently acquired wavefield $\upi$ by $N$ measurements. In terms of the decorrelation coefficient, 
		\begin{equation}
 \Kroll (t_s) = 1- \dfrac { \wdwINT \upin \upiTau dt} { \Big[ \wdwINT \upinSQ dt \wdwINT \upiSQ dt \Big]^{1/2}}.
        \label{eq:RCC}
	    \end{equation}
	Provided the repeat measurement period $\RmeasT$, at which each $\upi$ wavefield is recorded is much smaller then the rate of change in the scattering coefficient $d g_0/ \RmeasT$, this formulation provides the flexibility of selecting a sufficiently small reference lag $N$ such that each correlation is able to better satisfy the assumptions of CWI and CWD.
	In terms of \cref{eq:K_the_homo} for a small $\RmeasT$ the measured $\Kroll$ and the associated $t_s$ provides a qualitative description of the rate-of-change $(\dot{g_{0}} = d g_{0_{pj-N}} / \RmeasT)$ in $g_0$ and $\delta v/v$ respectively throughout the monitoring period.  
\section{\label{sec:4} Experimental Setup}
\begin{figure}
\centering
\includegraphics[width=0.45\textwidth]{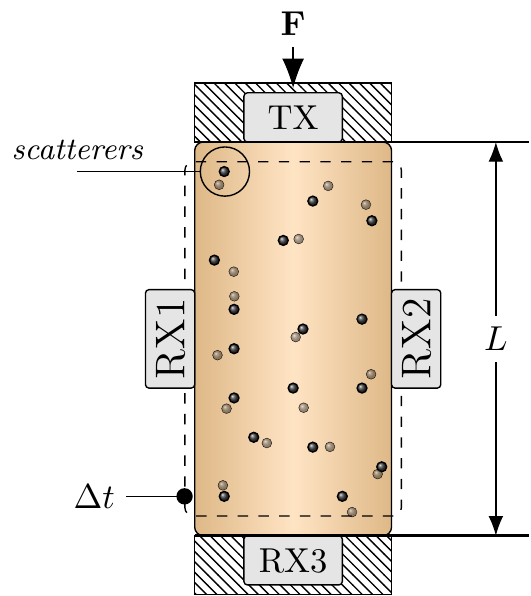}
\caption{\label{fig:Exp_setup}{Schematic representation of the UCS core test setup with active acoustic monitoring. An axially mounted \SI{}{\mega\hertz} source (TX) with one axial and two radial receives (RX) monitoring the changes in scattering properties every $\Delta t$ for increasing load F and axial deformation $\Delta L$.}}
\end{figure}

In order to experimentally analyse and compare the attributes of fixed and rolling reference forms of CWI and CWD for continuously evolving scattering properties, acoustic monitoring of Unconfined Compressive Strength (UCS) tests on laboratory scale core samples is made, see \cref{fig:Exp_setup}. For redundancy, one axial and two radial mounted transducers acting as receivers-RX, and one axially mounted source-TX transducer are attached to the surface of a core sample. Transducers with a peak operating frequency of $1$ or \SI{2.25}{\mega\hertz} are used depending on the experiment. In order to reduce the presence of noise, the stacking of 512 individual wavelets is made, resulting in a repeat measurement period $\RmeasT$ of \SI{15}{\second}. An axially increasing force $F$ is applied to each core sample controlled by a constant axial strain rate $(\Delta L/ Ldt)$ between 1.32 - \SI{4e-6}{\per\second}. The applied force $F$ and total axial displacement are recorded every half second throughout monitoring up until the sample experiences dynamic failure. The lithologies tested are Bentheimer sandstone \citep{Peksa2015} which has well sorted grain sizes (180-\SI{320}{\micro\meter}) and high porosity (21-\SI{27}{\percent}) and a poorly sorted (30-\SI{800}{\micro\meter}), zero primary porosity granite sourced from Benin. A summary of the three core sample UCS experiments is provided in \cref{tbl:UCSsummary} along with the abbreviated sample names employed in this work. 

\begin{table*}[t]
\centering
\caption{Summary of the UCS tests core sample Length and Diameter, lithology,  displacement rate, and acoustic source frequency.}	
\label{tbl:UCSsummary}
\begin{tabular}{P{1cm}P{3cm}P{5cm}P{2cm}P{2cm}}
 &L/D $[mm]$&Lithology& $\Delta L /Ldt \ $[\SI{}{\per\second}]& Src. $[\SI{}{\mega\hertz}]$\\
\hline
BNT1  & 75.01/39.75 & Bentheimer sandstone &\num{4.0e-6}  & 1.0 \\
BNT2  & 75.11/30.00 & Bentheimer sandstone &\num{4.0e-6} & 1.0 \\
GRA    & 75.70/29.80  & Benin granite                   & \num{1.3e-6} &  2.25 \\
\end{tabular}
\end{table*}

\section{\label{sec:5} CWI and CWD Monitoring of an UCS test}
\begin{figure*}[ht]
\centering
\baselineskip=12pt 
\includegraphics[width=.8\textwidth]{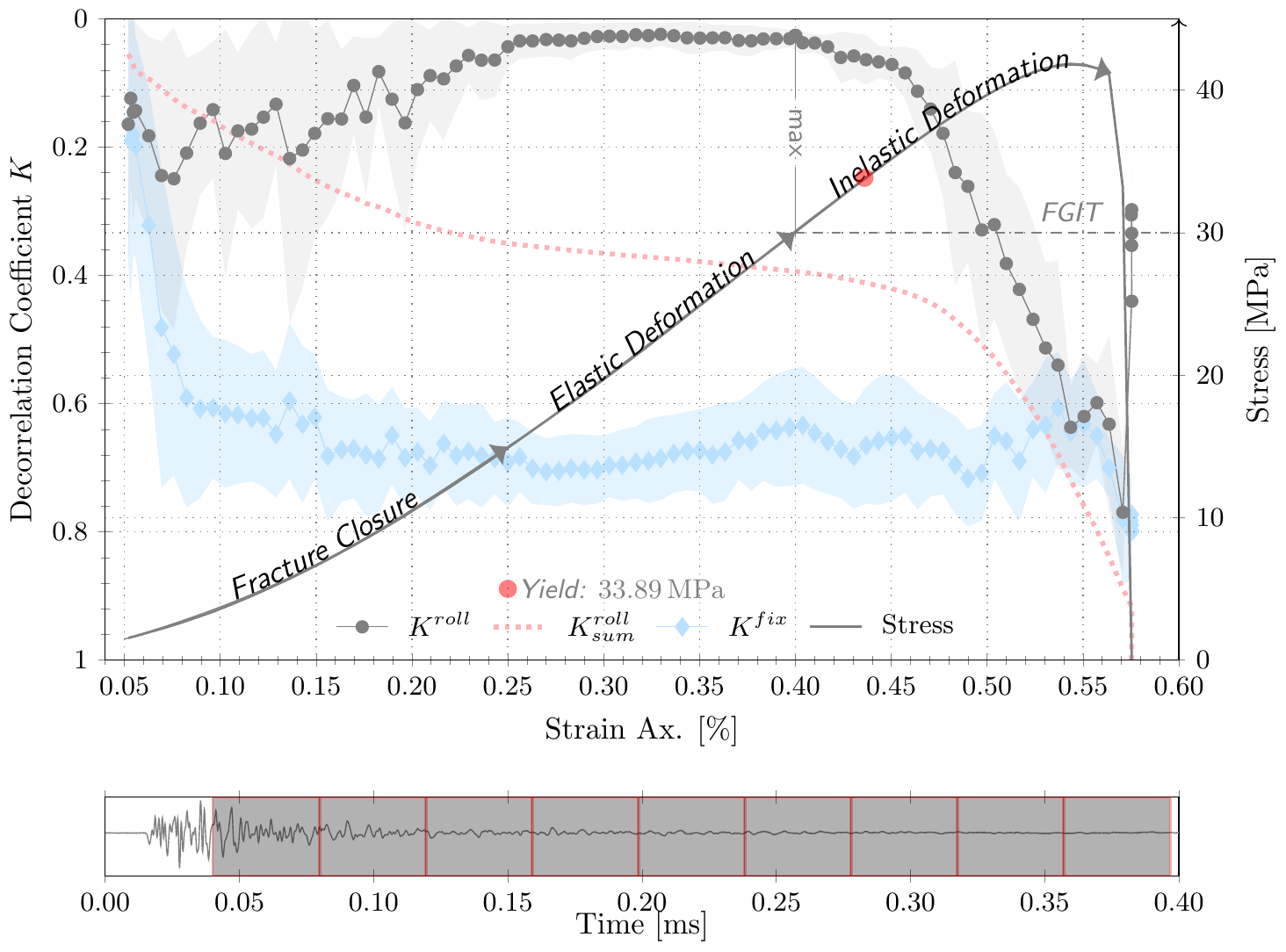}
  \caption{\label{fig:fixedrollBent} {CWD monitoring (receiver 2) of BNT1. The mean fixed reference decorrelation coefficient $\Kfix$ is compared with the rolling reference equivalent $\Kroll$ and its cumulative sum $\KrollInt$, calculated from the 9 independent correlation windows. The $\Kroll$ trend is generated with a rolling lag equal to the repeat measurement period of \SI{15}{\second} (i.e. $N=1$), from which the FGIT is identified at \SI{30}{\mega\pascal}. The shaded area represents the standard deviation about the mean decorrelation as calculated from the \SI{0.04}{\milli\second} correlation windows shown over a single waveform.}}
\end{figure*}

Monitoring the evolution of a rock matrix towards dynamic failure with a wavelet at the mesoscopic scale \citep{Chunlin1998,Heap2008,Barnhoorn2010, Barnhoorn2018}, is characterised by the continuous evolution of its wavefield propagation properties. For Unconfined Compressive Strength tests this can be illustrated by considering the different stages of material deformation as shown in \cref{fig:fixedrollBent} for the BNT1 core sample. Firstly, the initial closure of any existing fractures or pore-space occurs. This is followed by the elastic deformation where a linear stiffening of the rock matrix is expected. These first two stages represent in terms of \cref{eq:g0} a general reduction in the total scattering cross-section $\TotCross$ and therefore $g_0$ of the medium, as the size and then impedance contrast of the fractures reduces. Finally, the onset of inelastic deformation marks the beginning of fracture growth/formation often termed the Fracture Initiation and Growth Threshold (FIGT) \citep{Heap2008}. The growth and addition of new fractures represents an irreversible increase in $\TotCross$ and number density of fractures $\NumberDen$ respectively. Considering these changes in material scattering properties over a typical UCS test, it becomes difficult to make the assumption of a phase dominated or weak change in $g_0$ necessary in order to apply either CWI or CWD for a fixed initial refernce. Focusing first on the Coda-Wave Decorrelation method, in \cref{fig:fixedrollBent} we compare the sensitivity of a fixed $\Kfix$ versus a rolling $\Kroll$ reference decorrelation coefficient to the stages of material deformation common to UCS tests.

The $\Kfix$ coefficient shows a rapid increase over the first \SI{0.025}{\percent} of axial strain, following by a more gradual slope towards \SI{0.7}{} at the onset of elastic deformation. For the remainder of the USC test $\Kfix$ shows little sensitivity to the ongoing deformation, with only a pronounced increase as dynamic failure occurs at \SI{0.56}{\percent} axial strain. This indicates initial fracture closure represents a large perturbation in $g_{0}$, though after this point $\Kfix$ provides little informative value other then the knowledge that some notable change occurred early on during the monitoring period. 
In comparison, $\Kroll$ which is related to the relative rate of change in $g_{0}$ as calculated by \cref{eq:RCC} shows clearly identifiable trends segmenting each of the defined regions of deformation.
	\begin{itemize}
		\item The initial \SI{0.025}{\percent} of axial strain results in an increase in the rate of change of decorrelation $\Kroll$. This indicates the initial deformation represents an increase in the rate of fracture closure, and therefore an increase in the rate of reduction in the materials scattering coefficient (i.e. curvature in $\ddot{g_{0}}<0$) is expected as the impedance contrast of inter or transgranular fractures is reduced.
		\item This is followed by a gradual reduction in $\Kroll$ indicating a reducing rate of fracture closure (i.e. curvature in $\ddot{g_{0}}>0$) as the rock matrix stiffens leading up to the beginning of elastic deformation. 
		\item The elastic deformation of the material is characterised by a constant, low $\Kroll$ and therefore $\dot{g_{0}}$, where $\ddot{g_{0}}=0$. Here a unit stiffening of the rock matrix results in a proportional change in the impedance contrast of fractures and therefore $g_{0}$.
		\item At the Fracture Initiation and Growth Threshold - FIGT inelastic deformation occurs, which results in an increase in $g_{0}$, where $\dot{g_0}$ and $\ddot{g_{0}}>0$. These changes are reflected by initially a gradual and then steep increase in $\Kroll$ all the way until dynamic failure. This initial gradual increase is possibly evidence of the sub inelastic region of stable fracture growth before unstable growth continues, noted in literature for crystalline and brittle rocks \citep{Bieniawski1967}.
	\end{itemize}
 In order to make a direct comparison with $\Kfix$ the scaled ($0-1$) average cumulative summation of each correlation windows decorrelation coefficient ($\KrollInt$) is also provided. A general agreement is evident between $\Kfix$ and $\KrollInt$ as both show an initial increase, stabilisation, and then final increase within the region of elastic deformation, however the latter clearly shows an improvement in its sensitivity to the underlying perturbations. Most notably, $\KrollInt$ reflects the linear region of elastic deformation as the applied constant strain rate results in a linear stiffening of the medium (i.e. a change in $\TotCross$) and the transition to the non-linear region of elastic fracturing. The smoothed appearance of $\KrollInt$ is a result of its calculation from the average of each independent correlation windows cumulatively summed decorrelation coefficient.

\begin{figure*}[ht]
\centering
\baselineskip=12pt 
\includegraphics[width=.8\textwidth]{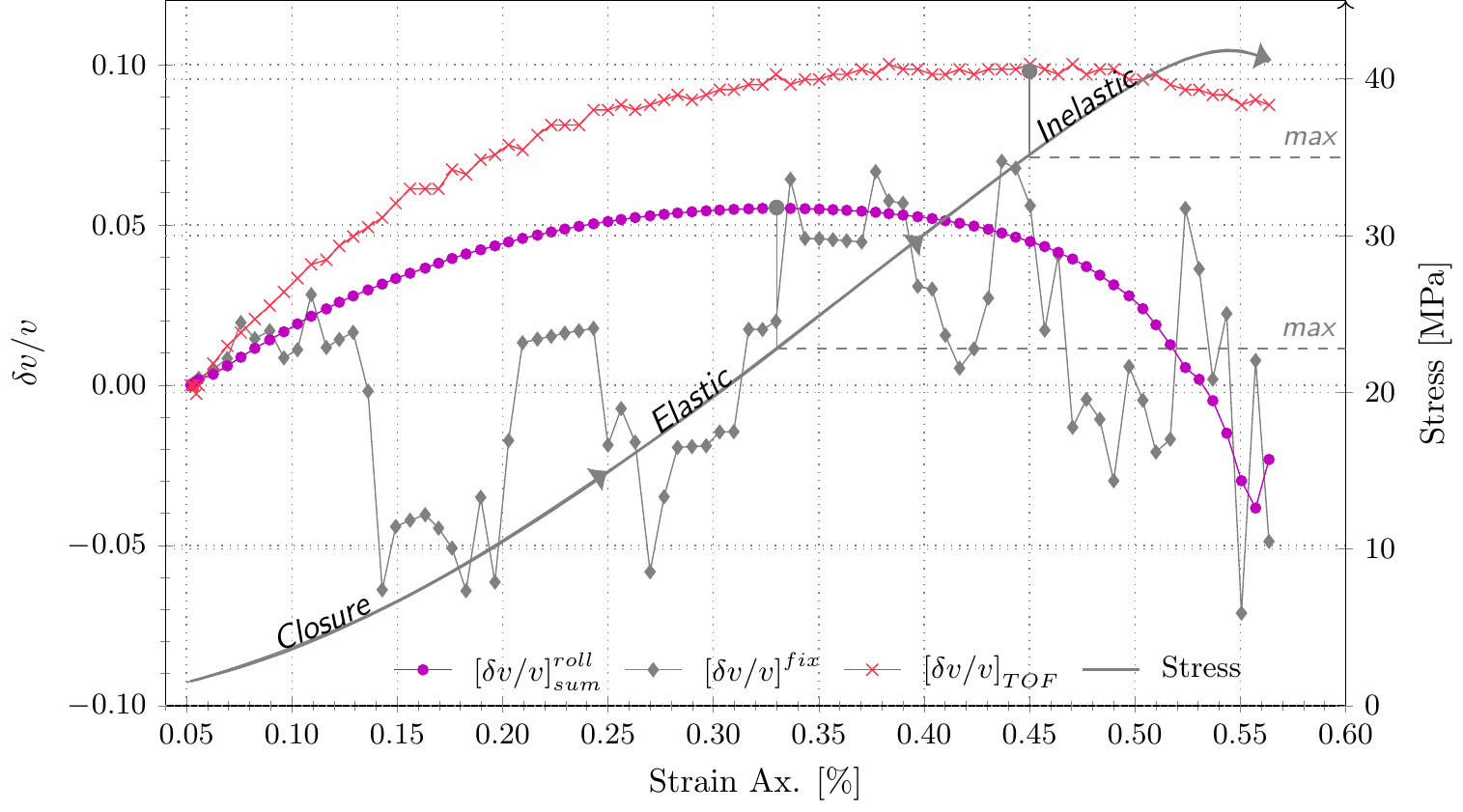}
  \caption{\label{fig:fixedrollTOFBent} { CWI monitoring (receiver RX2) of the BNT1 dataset. The mean fixed reference $\relVelFix$ relative velocity change is compared with the equivalent mean rolling reference cumulative summation $\relVelRoll$ (max: \SI{28.9}{\mega\pascal}) for the 9 windows shown in \cref{fig:fixedrollBent}. For comparison the Time Of Flight derived $\relVelTOF$ (max: \SI{35}{\mega\pascal}) is provided.}}
\end{figure*}

In making a similar comparison for Coda-Wave Interferometry monitoring of the BNT1 dataset, the differences in sensitivity between the fixed and rolling reference relative velocity changes as determined by \cref{eq:relVel} are compared, see \cref{fig:fixedrollTOFBent}. The fixed reference $\relVelFix$ shows an initial coherent increase in velocity inline with the Time Of Flight changes $[\delta v/v]^{TOF}$ derived from the first arrivals, though after only the first \SI{0.12}{\percent} of axial strain its trend appears incoherent. In contrast, the rolling reference derived cumulative summation $\relVelRoll$ curve tracks the $\relVelTOF$ curve plateauing over the elastic region, before decaying with increasing inelastic deformation.  A maximum in $\relVelRoll$ occurs at \SI{22.8}{\mega\pascal} whereas the $\relVelTOF$ maximum occurs at \SI{35}{\mega\pascal}. The initial increase in relative velocity is explained by acoustoelastic theory, while any subsequent reduction in velocity occurs due to the formation of the first micro-cracks \citep{Selleck1998,Shah2010}. The earlier reduction in $\relVelRoll$ indicates the expected improvement in sensitivity of CWI to these inelastic changes \citep{Zhang2012}. The presented $\relVelRoll$ is the average relative velocity change determined from each of the 9 correlation windows show in \cref{fig:fixedrollBent}, and for the purpose of this study is only considered as a qualitative indicator of the materials velocity change. 

In summary when applying CWD the three stages of material deformation from initial fracture closure, elastic deformation, and fracture growth are all reflected in the $\Kroll$ and $\KrollInt$ trend lines, whereas it is difficult to find any clear segmentation in the $\Kfix$ curve. The application of a rolling reference CWI derived $\relVelRoll$, while not able to clearly segment each region of deformation does track the Time Of Flight derived changes with improved sensitivity to the onset of inelastic deformation, whereas the fixed reference $[\delta v/v]$ does not. 
\section{\label{sec:5} Precursory Identification of Material Yielding}
\begin{figure*}
\centering
\baselineskip=12pt 
\includegraphics[width=.8\textwidth]{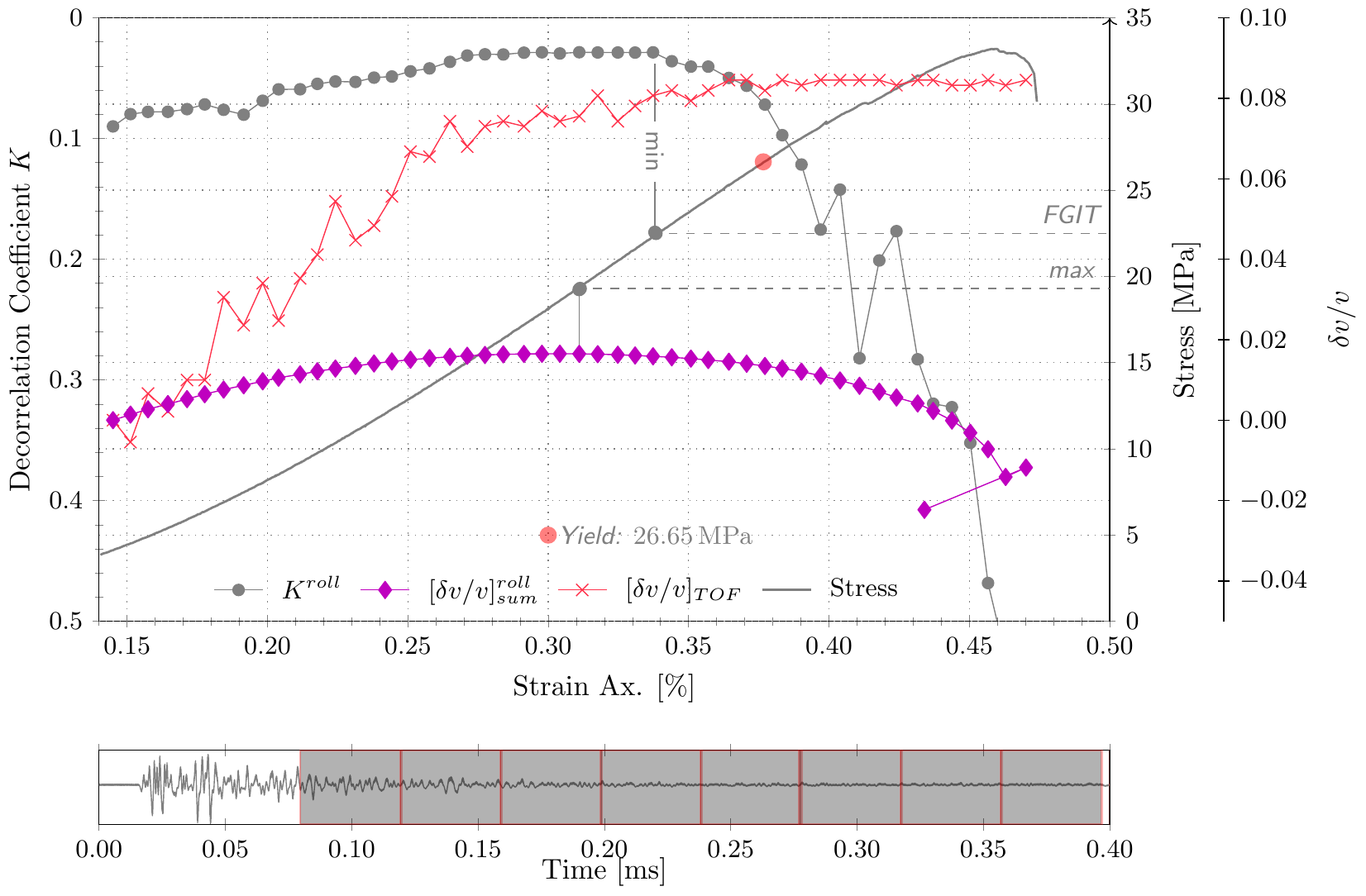}
  \caption{CWD and CWI monitoring of BNT2. The $\Kroll$ (FGIT: \SI{22.5}{\mega\pascal}) and $\relVelRoll$  (max: \SI{19.3}{\mega\pascal}) trends are generated with a rolling lag equal to the repeat measurement period of \SI{15}{\second} (i.e. $N=1$), calculated from the average of the 8 independent  \SI{0.04}{\milli\second} correlation windows. For comparison the Time Of Flight derived $\relVelTOF$ (max: \SI{34}{\mega\pascal}) is provided.}
  \label{fig:RCC_slice_BNT2MHz}
\end{figure*}

\begin{figure*}
	\centering
\baselineskip=12pt 
\includegraphics[width=.8\textwidth]{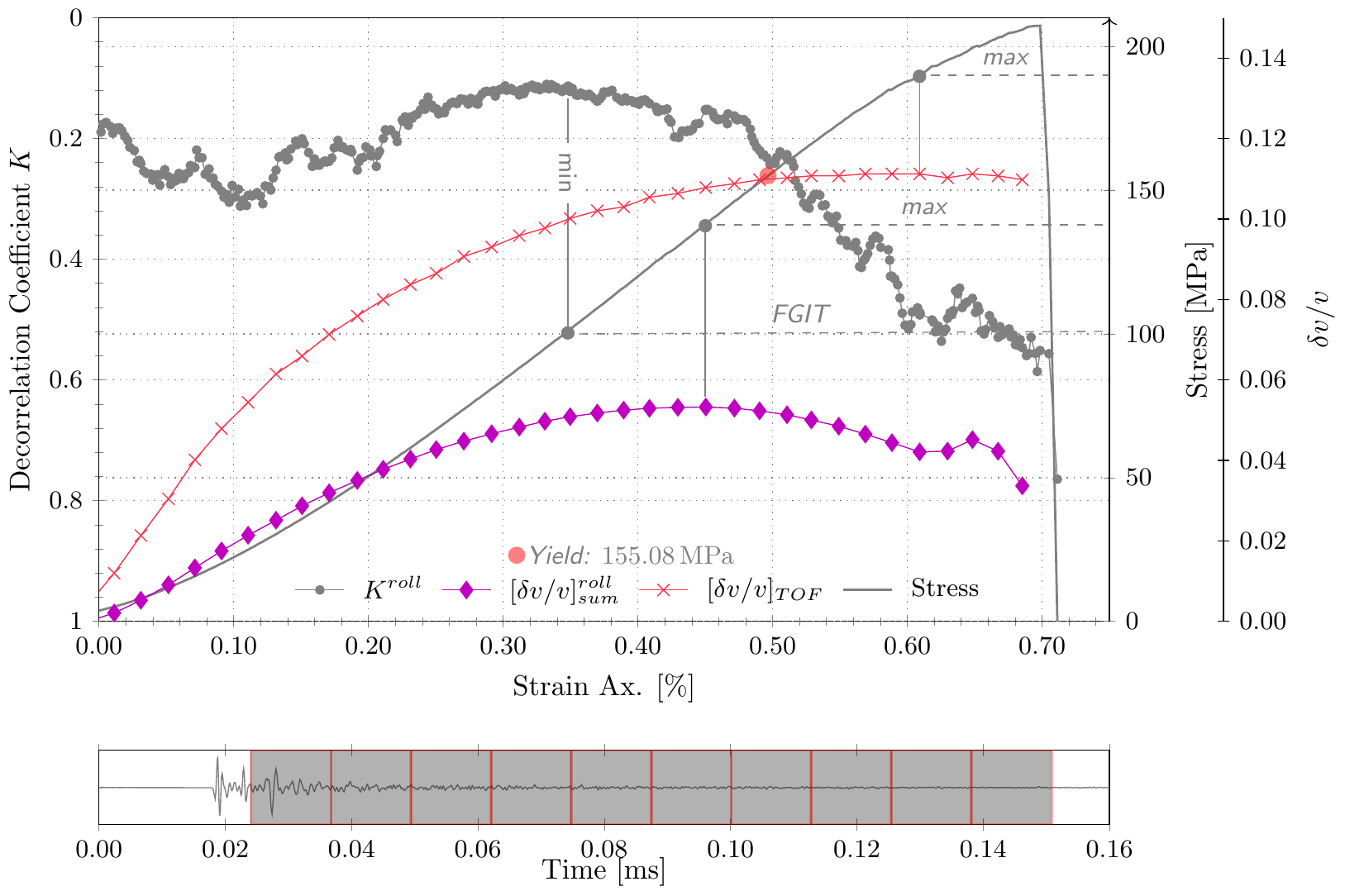}
  \caption{CWD and CWI monitoring of GRA. The $\Kroll$ (FGIT: \SI{101}{\mega\pascal}) trend is generated with a repeat measurement period of \SI{150} {\second} ($N=10$) from the average of the 10 independent, \SI{0.013}{\milli\second} correlation windows shown. The $\relVelRoll$ (max: \SI{138}{\mega\pascal})  cumulative summation is made for $N=1$. For comparison the Time Of Flight derived $\relVelTOF$ (max: \SI{190}{\mega\pascal}) is provided.}
  \label{fig:RCC_slice_GRA-2.25MHz}
\end{figure*}

\begin{figure*}
\centering
\baselineskip=12pt 
\includegraphics[width=.8\textwidth]{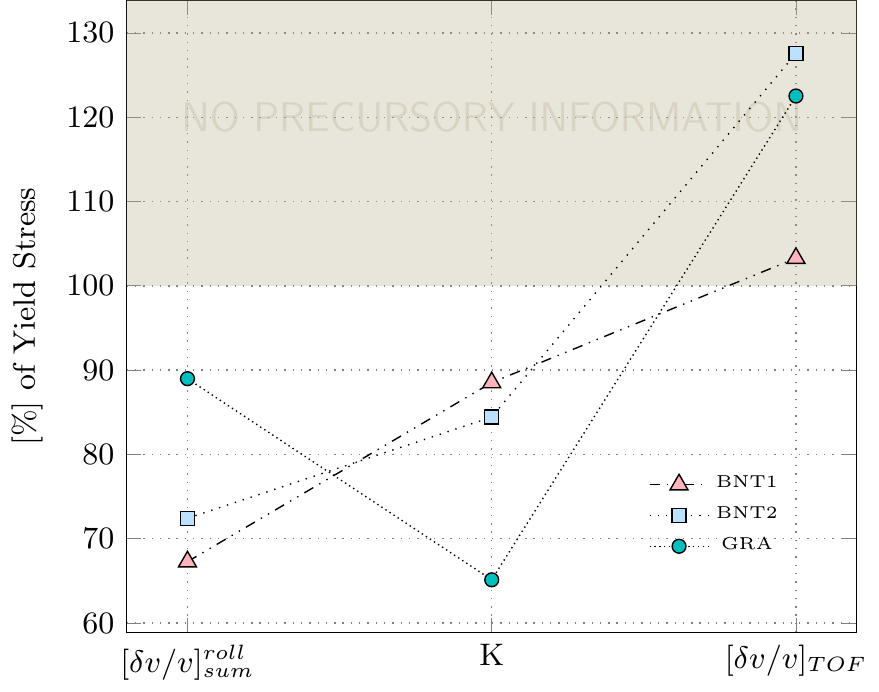}
  \caption{(color online) Comparison of the precursory indicators to the Yield stress, determined from the rolling reference integrated relative velocity changes $\relVelRoll$, the rolling reference decorrelation coefficient $\Kroll$, and the TOF derived relative velocity changes $\relVelTOF$.}
  		\label{fig:PrecurPower}
\end{figure*}

Based on the presented ability of both $\Kroll$ and $\relVelRoll$ to monitor deformation of a rock matrix, we will now assess both methods sensitivity to the onset of inelastic deformation.  The yield point of a material is often defined as the transition from predominantly stable fracture growth to predominantly unstable fracture growth \citep{Bieniawski1967, Mogi2007}, while others define it as the point the first micro-fractures are formed \citep{Elliott1986, Paterson2005}. There is however, general consensus that the yield point can be identified on a stress/strain diagram as the onset of non-linear behaviour which follows a linear elastic region. This parameter is of critical importance in the prediction of dynamic material failure as it signifies that permanent material deformation has begun.  Typically, the yield point from a UCS test would be determined by hand, though in order to remove some of the ambiguity surrounding this, an automated search is made based on the method described in the Appendix. On the basis of this yield point the precursory/subtle detection capabilities of the CWD and CWI methods can be assessed. \newline
	For this purpose two additional UCS tests are made, on a repeat Bentheimer (\cref{fig:RCC_slice_BNT2MHz} - BNT2) and Benin granite (\cref{fig:RCC_slice_GRA-2.25MHz} - GRA) core sample. As with BNT1, the indication of inelastic deformation (FIGT) is identified at the end of the trough in $\Kroll$ as the rate of change in $g_0$ increases. Similarly, the coda and first arrival derived relative velocity change inelastic indicators are identified at the onset of a reduction in velocity as the acoustoelastic effect \citep{Lee2014} working to increase the velocity is overcome by the formation of the first micro-fractures \citep{Zhang2012}.  In order to make a comparison between the different lithologies and datasets, each precursory indicator is quoted in terms of the percentage of the yield stress at which it is found, as presented in \cref{fig:PrecurPower}.
	
For the Bentheimer sandstone samples (BNT1 and 2) the CWI cumulative relative velocity changes $\relVelRoll$ indicate the onset of yielding at around \SI{70}{\percent} of the yield stress, where the CWD decorrelation coefficient $\Kroll$ begins to increase at around \SI{86}{\percent} of the yield stress. The GRA sample on the other-hand show  $\Kroll$ as providing the earliest indication of inelastic deformation at \SI{66}{\percent} in comparison to $\relVelRoll$ at  \SI{89}{\percent} of yield stress. This may be a result of the differences in fracturing behaviour between the well sorted quartz-rich, porous grains of the Bentheimer sandstone and the poorly sorted non-porous grains of the Benin granite resulting in reduced sensitivity of the velocity change at the onset of fracturing.

In all cases the reference $\relVelTOF$ show no precursory indicative power to material yielding with the onset of decay occurring after the stress-strain identified yield point. The lack of sensitivity of Time-Of-Flight changes to the classically defined yield point which was also noted in the work by \citet{Barnhoorn2018}. Refer to the Appendix for a description of the method applied to determine the yield point. 	

\section{\label{sec:7} Discussion and Conclusions}
In this study, we have shown that by applying a rolling reference waveform in both the CWI and CWD form, one is able to monitor the evolution of a material's relative velocity change and scattering coefficient where a large inelastic perturbation in its scattering properties occurs. Furthermore, we demonstrate that both approaches provide precursory indication of material yielding as determined from stress-strain measurements. While these benefits have direct applications in terms of providing a better understanding of rock properties in the laboratory, it is expected that such a processing approach will enable a wider range of field scale monitoring applications. 

In the structural concrete bending tests by \citet{Zhang2016a} which applied fixed reference CWI and CWD and the work by \citet{Larose2015} where only CWD was applied the authors noted the increase in error of both relative velocity and the decorrelation coefficient as inelastic deformation progressed. While this can be easily overcome by the manual selection of a new reference point during monitoring, we suggest that where the long term, continuously monitoring of both gradual, and sudden inelastic changes is required a rolling reference is the preferred choice. In terms of structural applications, this allows for the gradual long term monitoring of micro-cracking which occur throughout the life of concrete while maintaining sensitivity as an early warning to material failure. Furthermore, to the authors knowledge all application of CWD where notable inelastic deformation occurs have been under a bending force resulting in localised micro-cracks \citep{Paterson2005} where nucleation occurs within a region of tensile stress. As supported by the current study where a homogenous compressional stress field is applied, the issue of rapid decorrelation is likely to become more immediately evident where a more homogeneously distributed fracture network forms.

An active source landslide or rock slide is an example of a near surface environmental geophysics application, where long term monitoring for the purpose of early warning detection to dynamic failure is required. Passive monitoring has been shown to be sensitive to a continuous velocity change, prior to a fluid slippage event \citep{Mainsant2012}, where a rapid reduction in rigidity within a liquefaction region was identified as the mechanism for initial dislocation.
Pore-pressure driven reservoir compaction is a process which is by definition controlled at the grain scale \citep{Hol2015} with the first large scale induced seismicity events often along existing faults occurring decades after the beginning of production \citep{Bourne2014,VanThienen-Visser2015}. An in borehole active source continuous monitoring setup \citep{Zoback2011}, will perhaps be able to identify both the gradual inelastic intergranular compaction \citep{Pijnenburg2018} as well as the onset of larger scale transgranular cracking which results in surface subsidence \citep{Pratt1926a,Sharp1995,Pijpers2016}. 
Where long term continuous monitoring is required it is expected that rolling reference processing will help fulfill some of the criteria for Early Warning System \citep{Staehli2015,Michoud2013a,Nations}, particularly in terms of the robustness, and autonomy of the system.

In the context of this study, the ability to determine the subtle onset of material yielding was possible with only the knowledge that loading occurs at a constant strain rate, and in the absence of direct stress and strain measurements. Provided the employed source produces wavelengths sensitive to the perturbations in scattering, sensitivity to the mentioned structural-health, landslide, and compaction related monitoring is expected. While the feasibility of such applications requires additional investigation they can all be generalised as monitoring scenarios where both high rate-of-change and long-term transient changes in material scattering properties are of concern. It is therefore suggested that with an approximate knowledge of the underlying direction and type of perturbation, rolling reference coda based monitoring can be a useful addition to the mentioned Early Warning Systems as well as the long term monitoring of a materials scattering properties.
\section*{acknowledgments}
We would like to thank Christian Reinicke, Cornelis Weemstra, Deyan Draganov, Kees Wapenaar, Lisanne Douma and Richard Bakker for their fruitful discussion throughout this work. Furthermore, the comments made by two anonymous reviewers helped to improve the brevity, clarity and completeness of the work. A special thanks goes to the laboratory support staff at TUDelft, and specifically Karel Heller for his assistance in setting up an automated ultrasonic acquisition system for our experimental work. The recorded data used in this study can be found at https://data.4tu.nl under DOI uuid:e3071a50-7310-4609-95a1-6f8e69e556e2.
\appendix

	\section{\label{APP:yield}Yield Point Identification}
	The yield point of a stress($\sigma$)-strain($\epsilon$) curve is determined by a search for the most linear region. This is achieved by fitting $q$ lines of length $m$ to the trend as described by,
		 \begin{equation}
	    	\sigma_{i}^{fit} = \beta_{0} + \beta_{1} \epsilon_{i} \biggr\rvert _{i=1}^{m},
	    	\label{eq:YieldFit}
	    \end{equation}
	    and assessing the most linear on the conditions that the slope $\beta_{1}$ is at a maximum and the fitting error at the end of the line is less than $2 \%$.
	    \begin{align}
	    	\beta_{q,1} > \beta_{q-1,1} \\
	    	\Delta(\sigma_{q,m}^{fit}, \sigma_{q,m}) < 2 \%.
	    	\label{eq:YieldCond}
	    \end{align}
	    Here $m$ is selected to be approximately half the length of the linear elastic region in terms of sample points. The yield point is then defined at the end of the fitted line with maximum slope $\beta_{1}$ and fitting error $<2\%$.
	

\begin{thebibliography}{}

\bibitem[Aki and Chouet, 1975]{Aki1975}
Aki, K. and Chouet, B. (1975).
\newblock {Origin of coda waves: Source, attenuation, and scattering effects}.
\newblock {\em Journal of Geophysical Research}, 80(23):3322--3342.

\bibitem[Barnhoorn et~al., 2010]{Barnhoorn2010}
Barnhoorn, A., Cox, S.~F., Robinson, D.~J., and Senden, T. (2010).
\newblock {Stress- and fluid-driven failure during fracture array growth:
  Implications for coupled deformation and fluid flow in the crust}.
\newblock {\em Geology}, 38(9):779--782.

\bibitem[Barnhoorn et~al., 2018]{Barnhoorn2018}
Barnhoorn, A., Verheij, J., Frehner, M., Zhubayev, A., and Houben, M. (2018).
\newblock {Experimental identification of the transition from elasticity to
  inelasticity from ultrasonic attenuation analyses}.
\newblock {\em GEOPHYSICS}, 83(4):MR221--MR229.

\bibitem[Bieniawski, 1967]{Bieniawski1967}
Bieniawski, Z.~T. (1967).
\newblock {Mechanism of brittle fracture of rock; Part I-theory of the fracture
  process}.
\newblock {\em International Journal of Rock Mechanics and Mining Sciences},
  4:395--406.

\bibitem[Bourne et~al., 2014]{Bourne2014}
Bourne, S.~J., Oates, S.~J., van Elk, J., and Doornhof, D. (2014).
\newblock {A seismological model for earthquakes induced by fluid extraction
  from a subsurface reservoir}.
\newblock {\em Journal of Geophysical Research: Solid Earth},
  119(12):8991--9015.

\bibitem[Brownjohn, 2007]{Brownjohn2007}
Brownjohn, J. M.~W. (2007).
\newblock {Structural health monitoring of civil infrastructure}.
\newblock {\em Philosophical Transactions of the Royal Society A: Mathematical,
  Physical and Engineering Sciences}, 365(1851):589--622.

\bibitem[Elliott and Brown, 1986]{Elliott1986}
Elliott, G.~M. and Brown, E.~T. (1986).
\newblock {Further development of a plasticity approach to yield in porous
  rock}.
\newblock {\em International Journal of Rock Mechanics and Mining Sciences
  and}, 23(2):151--156.

\bibitem[Gr{\^{e}}t, 2005]{Gret2005}
Gr{\^{e}}t, A. (2005).
\newblock {Monitoring rapid temporal change in a volcano with coda wave
  interferometry}.
\newblock {\em Geophysical Research Letters}, 32(6):L06304.

\bibitem[Gr{\^{e}}t et~al., 2006]{Gret2006}
Gr{\^{e}}t, A., Snieder, R., and Scales, J. (2006).
\newblock {Time-lapse monitoring of rock properties with coda wave
  interferometry}.
\newblock {\em Journal of Geophysical Research: Solid Earth}, 111(3):1--11.

\bibitem[Heap and Faulkner, 2008]{Heap2008}
Heap, M. and Faulkner, D. (2008).
\newblock {Quantifying the evolution of static elastic properties as
  crystalline rock approaches failure}.
\newblock {\em International Journal of Rock Mechanics and Mining Sciences},
  45(4):564--573.

\bibitem[Hol et~al., 2015]{Hol2015}
Hol, S., Mossop, A., van~der Linden, A., Zuiderwijk, P., and Makurat, A.
  (2015).
\newblock {Long-term compaction behavior of Permian sandstones - An
  investigation into the mechanisms of subsidence in the Dutch Wadden Sea}.
\newblock {\em Arma}, 15-618.

\bibitem[Larose et~al., 2015]{Larose2015}
Larose, E., Obermann, A., Digulescu, A., Plan{\`{e}}s, T., Chaix, J.-F.,
  Mazerolle, F., and Moreau, G. (2015).
\newblock {Locating and characterizing a crack in concrete with diffuse
  ultrasound: A four-point bending test}.
\newblock {\em The Journal of the Acoustical Society of America},
  138(1):232--241.

\bibitem[Larose et~al., 2010]{Larose2010}
Larose, E., Planes, T., Rossetto, V., and Margerin, L. (2010).
\newblock {Locating a small change in a multiple scattering environment}.
\newblock {\em Applied Physics Letters}, 96(20).

\bibitem[Li et~al., 1998]{Chunlin1998}
Li, C., Prikryl, R., and Nordlund, E. (1998).
\newblock {The stress-strain behaviour of rock material related to fracture
  under compression}.
\newblock {\em Engineering Geology}, 49(3-4):293--302.

\bibitem[Mainsant et~al., 2012]{Mainsant2012}
Mainsant, G., Larose, E., Br{\"{o}}nnimann, C., Jongmans, D., Michoud, C., and
  Jaboyedoff, M. (2012).
\newblock {Ambient seismic noise monitoring of a clay landslide: Toward failure
  prediction}.
\newblock {\em Journal of Geophysical Research: Earth Surface},
  117(F1):n/a--n/a.

\bibitem[Matsumoto et~al., 2001]{Matsumoto2001a}
Matsumoto, S., Obara, K., Yoshimoto, K., Saito, T., Ito, A., and Hasegawa, A.
  (2001).
\newblock {Temporal change in P -wave scatterer distribution associated with
  the M 6.1 earthquake near Iwate volcano, northeastern Japan}.
\newblock {\em Geophysical Journal International}, 145(1):48--58.

\bibitem[Michoud et~al., 2013]{Michoud2013a}
Michoud, C., Bazin, S., Blikra, L.~H., Derron, M.~H., and Jaboyedoff, M.
  (2013).
\newblock {Experiences from site-specific landslide early warning systems}.
\newblock {\em Natural Hazards and Earth System Sciences}, 13(10):2659--2673.

\bibitem[Mogi, 2007]{Mogi2007}
Mogi, K. (2007).
\newblock {\em {Experimental rock mechanics}}.
\newblock CRC Press.

\bibitem[Obermann et~al., 2013a]{Obermann2013a}
Obermann, A., Plan{\`{e}}s, T., Larose, E., and Campillo, M. (2013a).
\newblock {Imaging preeruptive and coeruptive structural and mechanical changes
  of a volcano with ambient seismic noise}.
\newblock {\em Journal of Geophysical Research: Solid Earth},
  118(12):6285--6294.

\bibitem[Obermann et~al., 2013b]{Obermann2013}
Obermann, A., Planes, T., Larose, E., Sens-Sch{\"{o}}nfelder, C., and Campillo,
  M. (2013b).
\newblock {Depth sensitivity of seismic coda waves to velocity perturbations in
  an elastic heterogeneous medium}.
\newblock {\em Geophysical Journal International}, 194:372--382.

\bibitem[Paasschens, 1997]{Paasschens1997}
Paasschens, J. (1997).
\newblock {Solution of the time-dependent Boltzmann equation}.
\newblock {\em Physical Review E}, 56(1):1135--1141.

\bibitem[Pacheco and Snieder, 2005]{Pacheco2005}
Pacheco, C. and Snieder, R. (2005).
\newblock {Time-lapse travel time change of multiply scattered acoustic waves}.
\newblock {\em The Journal of the Acoustical Society of America}, 118(3):1300.

\bibitem[Page et~al., 2000]{Page2000}
Page, J.~H., Cowan, M.~L., and Weitz, D.~A. (2000).
\newblock {Diffusing acoustic wave spectroscopy of fluidized suspensions}.
\newblock {\em Physica B: Condensed Matter}, 279(1-3):130--133.

\bibitem[Paterson and Wong, 2005]{Paterson2005}
Paterson, M.~S. and Wong, T.~F. (2005).
\newblock {\em {Experimental Rock Deformation — The Brittle Field}}.
\newblock Springer-Verlag, Berlin/Heidelberg.

\bibitem[Peksa et~al., 2015]{Peksa2015}
Peksa, A.~E., Wolf, K. H.~A., and Zitha, P.~L. (2015).
\newblock {Bentheimer sandstone revisited for experimental purposes}.
\newblock {\em Marine and Petroleum Geology}, 67:701--719.

\bibitem[Pijnenburg et~al., 2018]{Pijnenburg2018}
Pijnenburg, R. P.~J., Verberne, B.~A., Hangx, S.~J., and Spiers, C.~J. (2018).
\newblock {Deformation Behavior of Sandstones From the Seismogenic Groningen
  Gas Field: Role of Inelastic Versus Elastic Mechanisms}.
\newblock {\em Journal of Geophysical Research: Solid Earth},
  123(7):5532--5558.

\bibitem[Pijpers and {Van der Laan}, 2016]{Pijpers2016}
Pijpers, F. and {Van der Laan}, D.~J. (2016).
\newblock {Trend changes in ground subsidence in Groningen}.
\newblock (May).

\bibitem[Pine et~al., 1990]{Pine1990}
Pine, D., Weitz, D., Zhu, J., and Herbolzheimer, E. (1990).
\newblock {Diffusing-wave spectroscopy: dynamic light scattering in the
  multiple scattering limit}.
\newblock {\em Journal de Physique}, 51(18):2101--2127.

\bibitem[Plan{\`{e}}s, 2013]{Planes2013a}
Plan{\`{e}}s, T. (2013).
\newblock {\em {Imagerie de chargements locaux en regime de diffusion
  multiple}}.
\newblock PhD thesis, Universit{\'{e}} de Grenoble.

\bibitem[Plan{\`{e}}s et~al., 2013]{Planes2013c}
Plan{\`{e}}s, T., Larose, E., Rossetto, V., and Margerin, L. (2013).
\newblock {LOCADIFF: Locating a weak change with diffuse ultrasound}.
\newblock 405(2013):405--411.

\bibitem[Plan{\`{e}}s et~al., 2015]{Planes2015}
Plan{\`{e}}s, T., Larose, E., Rossetto, V., and Margerin, L. (2015).
\newblock {Imaging multiple local changes in heterogeneous media with diffuse
  waves}.
\newblock {\em The Journal of the Acoustical Society of America},
  137(2):660--667.

\bibitem[Poupinet et~al., 1984]{Poupinet1984}
Poupinet, G., Ellsworth, W.~L., and Frechet, J. (1984).
\newblock {Monitoring velocity variations in the crust using earthquake
  doublets: An application to the Calaveras Fault, California}.
\newblock {\em Journal of Geophysical Research}, 89(B7):5719--5731.

\bibitem[Pratt and Johnson, 1926]{Pratt1926a}
Pratt, W.~E. and Johnson, D.~W. (1926).
\newblock {Local Subsidence of the Goose Creek Oil Field}.
\newblock {\em The Journal of Geology}.

\bibitem[Ratdomopurbo and Poupinet, 1995]{Ratdomopurbo1995}
Ratdomopurbo, A. and Poupinet, G. (1995).
\newblock {Monitoring a temporal change of seismic velocity in a volcano:
  Application to the 1992 eruption of Mt. Merapi (Indonesia)}.
\newblock {\em Geophysical Research Letters}, 22(7):775--788.

\bibitem[Rossetto et~al., 2011]{Rossetto2011}
Rossetto, V., Margerin, L., Plan{\`{e}}s, T., and Larose, E. (2011).
\newblock {Locating a weak change using diffuse waves: Theoretical approach and
  inversion procedure}.
\newblock {\em Journal of Applied Physics}, 109(3):034903.

\bibitem[Selleck et~al., 1998]{Selleck1998}
Selleck, S.~F., Landis, E.~N., Peterson, M.~L., Shah, S.~P., and Achenbach,
  J.~D. (1998).
\newblock {Ultrasonic investigation of concrete with distributed damage}.
\newblock {\em ACI Materials Journal}, 95(1):27--36.

\bibitem[Sens-Sch{\"{o}}nfelder and Larose, 2008]{Sens-Schonfelder2008}
Sens-Sch{\"{o}}nfelder, C. and Larose, E. (2008).
\newblock {Temporal changes in the lunar soil from correlation of diffuse
  vibrations}.
\newblock {\em Physical Review E - Statistical, Nonlinear, and Soft Matter
  Physics}, 78(4):1--4.

\bibitem[Sens-Sch{\"{o}}nfelder and Wegler, 2006]{Sens-Schonfelder2006}
Sens-Sch{\"{o}}nfelder, C. and Wegler, U. (2006).
\newblock {Passive image interferometry and seasonal variations of seismic
  velocities at Merapi Volcano, Indonesia}.
\newblock {\em Geophysical Research Letters}, 33(21):L21302.

\bibitem[Shah and Hirose, 2010]{Shah2010}
Shah, A.~A. and Hirose, S. (2010).
\newblock {Nonlinear Ultrasonic Investigation of Concrete Damaged under
  Uniaxial Compression Step Loading}.
\newblock {\em Journal of Materials in Civil Engineering}, 22(5):476--484.

\bibitem[Sharp and Hill, 1995]{Sharp1995}
Sharp, J.~M. and Hill, D.~W. (1995).
\newblock {Land subsidence along the northeastern Texas Gulf coast: Effects of
  deep hydrocarbon production}.
\newblock {\em Environmental Geology}, 25(3):181--191.

\bibitem[Snieder, 2002]{Snieder2002b}
Snieder, R. (2002).
\newblock {Coda Wave Interferometry for Estimating Nonlinear Behavior in
  Seismic Velocity}.
\newblock {\em Science}, 295(5563):2253--2255.

\bibitem[Snieder et~al., 2002]{Snieder2002a}
Snieder, R., Douma, H., and Scales, J. (2002).
\newblock {Coda Wave Interferometry for Estimating Nonlinear Behavior in
  Seismic Velocity}.
\newblock {\em Science}, 295(March):2253--2255.

\bibitem[St{\"{a}}hler et~al., 2011]{Stahler2011}
St{\"{a}}hler, S.~C., Sens-Sch{\"{o}}nfelder, C., and Niederleithinger, E.
  (2011).
\newblock {Monitoring stress changes in a concrete bridge with coda wave
  interferometry.}
\newblock {\em The Journal of the Acoustical Society of America},
  129(4):1945--1952.

\bibitem[St{\"{a}}hli et~al., 2015]{Staehli2015}
St{\"{a}}hli, M., S{\"{a}}ttele, M., Huggel, C., McArdell, B.~W., Lehmann, P.,
  {Van Herwijnen}, A., Berne, A., Schleiss, M., Ferrari, A., Kos, A., Or, D.,
  and Springman, S.~M. (2015).
\newblock {Monitoring and prediction in early warning systems for rapid mass
  movements}.
\newblock Technical Report~4.

\bibitem[Toupin and Bernstein, 1961]{Lee2014}
Toupin, R.~A. and Bernstein, B. (1961).
\newblock {Sound Waves in Deformed Perfectly Elastic Materials. Acoustoelastic
  Effect}.
\newblock {\em The Journal of the Acoustical Society of America},
  33(2):216--225.

\bibitem[{UN/ISDR: Global}, 2006]{Nations}
{UN/ISDR: Global} (2006).
\newblock {Global Survey of Early Warning Systems}.

\bibitem[van Thienen-Visser and Breunese, 2015]{VanThienen-Visser2015}
van Thienen-Visser, K. and Breunese, J.~N. (2015).
\newblock {Induced seismicity of the Groningen gas field: History and recent
  developments}.
\newblock {\em The Leading Edge}, 34(6):664--671.

\bibitem[Zhang et~al., 2012]{Zhang2012}
Zhang, Y., Abraham, O., Grondin, F., Loukili, A., Tournat, V., Duff, A.~L.,
  Lascoup, B., and Durand, O. (2012).
\newblock {Study of stress-induced velocity variation in concrete under direct
  tensile force and monitoring of the damage level by using
  thermally-compensated Coda Wave Interferometry}.
\newblock {\em Ultrasonics}, 52(8):1038--1045.

\bibitem[Zhang et~al., 2016]{Zhang2016a}
Zhang, Y., Plan{\`{e}}s, T., Larose, E., Obermann, A., Rospars, C., and Moreau,
  G. (2016).
\newblock {Diffuse ultrasound monitoring of stress and damage development on a
  15-ton concrete beam}.
\newblock {\em The Journal of the Acoustical Society of America},
  139(4):1691--1701.

\bibitem[Zoback et~al., 2011]{Zoback2011}
Zoback, M., Hickman, S., and Ellsworth, W. (2011).
\newblock {Scientific Drilling Into the San Andreas Fault Zone{\&}{\#}151;An
  Overview of SAFOD{\&}{\#}146;s First Five Years}.
\newblock {\em Scientific Drilling}, (11, March 2011):14--28.

\end{thebibliography}

\end{document}